# Effect of varying mixture ratio of raw material powders on the thermoelectric properties of AlMgB$_{14}$-based materials prepared by spark plasma sintering.

Shota Miura[1], Hikaru Sasaki[1], Ken-ichi Takagi[2], and Takuya Fujima[1]*

1) Department of Mechanical Engineering, Tokyo City University, 1-28-1 Tamazutsumi, Setagaya, Tokyo, Japan.

2) Advanced Research Laboratory, Tokyo City University, 8-15-1 Todoroki, Setagaya, Tokyo, Japan.

Abstract

The thermoelectric properties of AlMgB$_{14}$-based materials prepared by spark plasma sintering were investigated. Al, Mg, and B powder were used as raw material powders. The raw powders were mixed using a V-shaped mixer, and then the mixture was sintered at 1673 K or 1773K. The mixture ratio of raw powders was varied around stoichiometric ratio of AlMgB$_{14}$. X-ray diffraction patterns of samples showed that all samples consist of AlMgB$_{14}$ and some by-products, MgAl$_2$O$_4$, B$_2$O and AlB$_{12}$. The Seebeck coefficient of the samples exhibited significant change depending on the varying mixture ratio. One sample exhibited a large negative value for the Seebeck coefficient (approximately -500 µV/K) over the temperature range from 573K to 1073 K, while others showed positive value (250-450 µV/K). Thus n-type AlMgB$_{14}$-based material has been realized simply by varying raw material ratio.

1. Introduction

As the consumption of fossil fuels is growing in developing countries in recent years, thermoelectric generation is gathering much attention as a way to recover the energy wasted in exhaust heat [1-3] because a thermoelectric generation system converts thermal energy into electric power [4]. A practical generation module consists of two types of thermoelectric materials, n-type and p-type, that are connected electrically in series and thermally in parallel [5]. The energetic efficiency of thermoelectric materials is evaluated as the dimensionless figure of merit $ZT = S^2\sigma T/\kappa$, where $S$ is the Seebeck coefficient, $\sigma$ is the electrical conductivity, $T$ and $\kappa$ are the temperature and the thermal conductivity, respectively [6]. The numerator in $Z$, $S^2\sigma$, which is called power factor, is used for evaluating the electrical power that can be generated under a certain temperature gap. Highly-effective thermoelectric materials require a large $S$, a high $\sigma$, and a low $\kappa$.

Higher borides are a promising candidate as a thermoelectric material for high temperature usage, because the typical higher borides have chemical stability at high temperature, large Seebeck coefficient, and low thermal conductivity [7-12]. Though a lot of p-type materials have been reported about this substance group, n-type is quite rare [12-14].

AlMgB$_{14}$ was reported to have an extremely large negative Seebeck coefficient (-6500 µV/K) [15], however subsequent studies have reported positive values [16,17]. Element doping can drastically change thermoelectric properties of base material as H. Kim and K. Kimura reported β-rhombohedral boron exhibited change of $S$ between positive and negative along with vanadium doping [18]. Hence, we assume that AlMgB$_{14}$ is also strongly affected by impurities on its thermoelectric properties, which enables negative $S$ under some impurity condition. In this paper, we explore the thermoelectric properties of

AlMgB$_{14}$-based material with raw-material ratio variation where excess materials to stoichiometric composition can work as impurities therein.

2. Experimental procedure

   Samples were prepared by spark plasma sintering (SPS) method. The raw powders were amorphous B (95.6% purity, H.C. Starck Ltd.), Mg (99.5% purity, Kojundo Chemical Lab. Co. Ltd.), and Al (99.9% purity, Kojundo Chemical Lab Co. Ltd.). The powders were mixed in a V-shaped mixer for 30 min. The mixture was then sintered under a pressure of 30 MPa. The raw material ratio and sintering temperature are shown in Table 1. The sintered bodies were characterized by X-ray diffraction (XRD; Bruker AXS Ltd., D8 Advance) with CuKα radiation. The Seebeck coefficient and electrical conductivity were measured at temperatures ranging from 573K to 1073K and 373K up to 1073K, respectively using a ZEM-1 (ULVAC-RIKO, Inc.).

Table. 1 Powder mixture ratio and Sintering temperature

| Sample number | Powder mixture ratio [molar ratio]<br>Al : Mg : B | Sintering temperature [K] |
|---|---|---|
| #1 | 0.75 : 0.78 : 14 | 1673 |
| #2 | 0.88 : 0.88 : 14 | |
| #3 | 1.00 : 1.00 : 14 | |
| #4 | 1.04 : 1.06 : 14 | |
| #5 | 0.88 : 0.88 : 14 | 1773 |
| #6 | 1.05 : 1.05 : 14 | |
| #7 | 1.04 : 1.06 : 14 | |

3. Results and discussion

3.1 Characterization

XRD patterns of the samples are shown in Fig. 1. The diffraction patterns for $AlMgB_{14}$ and $MgAl_2O_4$ were detected in all the samples. $MgAl_2O_4$ has been reported as a by-product of $AlMgB_{14}$; this formation is caused by oxygen contamination in raw powders [19-21]. The diffraction pattern for sample #1 exhibited the peaks for $AlMgB_{14}$, $MgAl_2O_4$, and $AlB_{12}$ as well as some indefinable peaks. The diffraction patterns for sample #3-4, and sample #6-7 indicated $B_2O$.

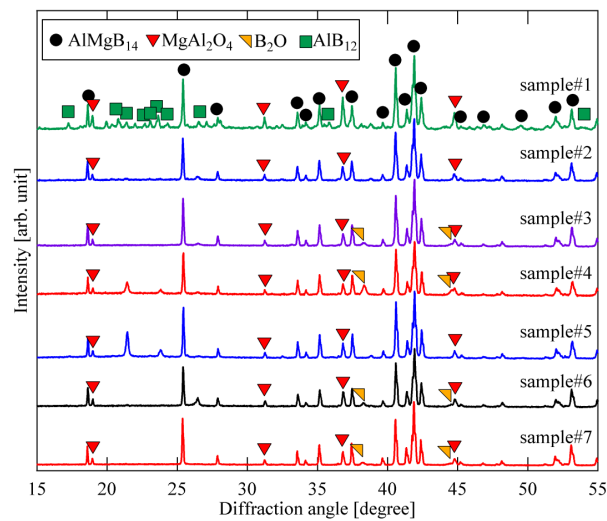

Fig. 1 X-ray diffraction patterns of sintered bodies

3.2 Thermoelectric properties

The temperature dependences of the Seebeck coefficient of the samples are shown in Fig. 2. Sample #1-#4 exhibits a positive Seebeck coefficient of 250-450 μV/K over the measured temperature range. For sample #5 and #6 that was sintered at 1773K, the value of the Seebeck coefficient was close to that of sample #1-#4. Sample #7 has a large negative

Seebeck coefficient ranging from 420 to 520 μV/K. This negative value has hardly been reported about the higher borides. Temperature dependence of sample #6 and #7 was quite different from each other though these two had similar powder mixture ratio.

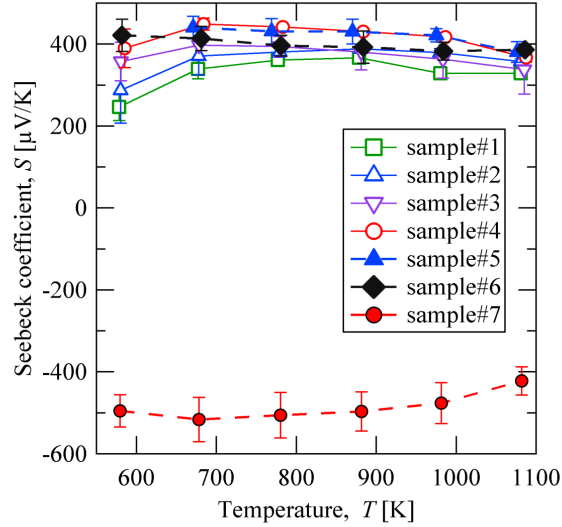

Fig. 2 Temperature dependence of the Seebeck coefficient

The temperature dependences of the electrical conductivity of the samples are in Fig. 3. All but sample #7 exhibited a similar shape of the curve over the measured temperature range.

Fig.4 shows the electrical conductivity in the logarithmic scale as a function of (a) $T^{-1}$, an Arrhenius plot, and (b) $T^{-1/4}$. The latter is based on the Mott's law of variable range hopping (VRH) conduction [22]:

$$\sigma_{VRH} = \sigma_0 \exp\left\{-\left(\frac{T_0}{T}\right)^{\frac{1}{4}}\right\}, T_0 = \frac{60\alpha^3}{\pi N(E_F)k_B}$$

where $\sigma_0$ is a constant, $T_0$ is the Mott temperature, $\alpha$, $N(E)$, $E_F$ and $k_B$ are the inverse of the localization length, the density of state, the Fermi energy and the Boltzmann constant, respectively. Though the temperature dependence of sample #1-#6 are linear in higher temperature range in both figure 4 (a) and (b), the difference among these conductions is small. Thus, it is difficult to determine the conduction type of those. By contrast, the temperature dependence of sample #7 is not linear, which indicates the conduction mechanism in these materials is neither a band conduction [23] nor VRH over the investigated temperature range.

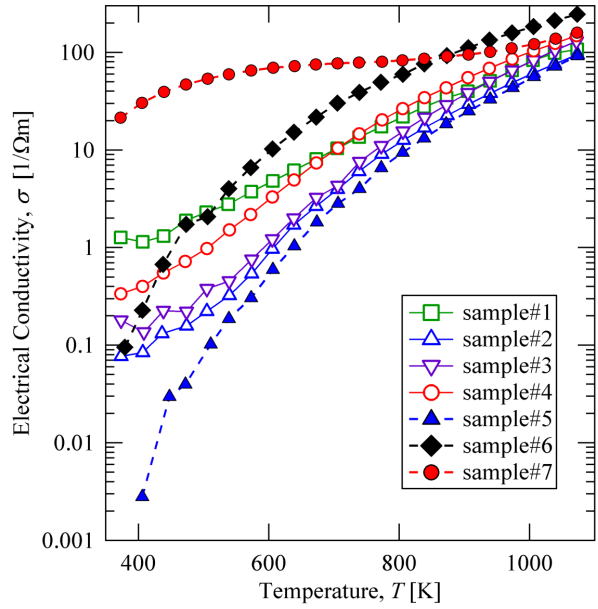

Fig. 3 Temperature dependence of the Electrical conductivity.

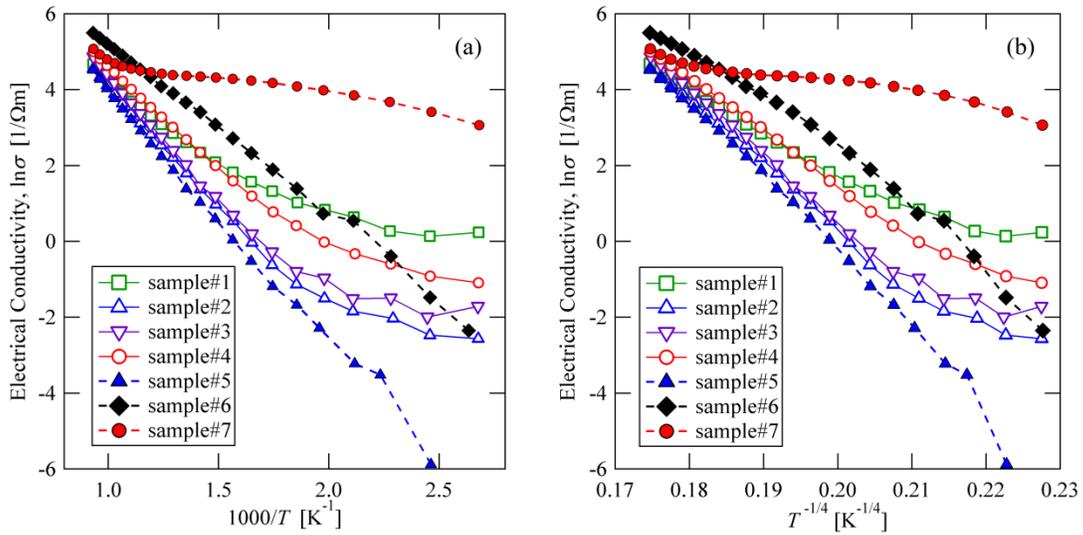

Fig. 4 Temperature dependence of the Electrical conductivity. Electrical conductivity is plotted against $T^{-1}$ in accordance with band conduction in (a) and $T^{-1/4}$ in accordance with Mott's law of variable range hopping conduction in (b).

4. Conclusion

We investigated thermoelectric properties of the AlMgB$_{14}$-based materials prepared by SPS with varying the mixture ratio of raw material powders, Al, Mg and B. Not clearly affecting the component of the sintered bodies, the variation significantly changed their thermoelectric properties such as Seebeck coefficient. The sample with raw material powder mixture ratio of Al:Mg:B =1.04:1.06:14 sintered at 1773K showed a large negative Seebeck coefficient ranging from -420 to -520 μV/K over a wide temperature range from 573K to over 1073K. The others exhibited positive Seebeck coefficient in all temperature range. Thus n-type AlMgB$_{14}$-based material has been realized simply by varying raw material ratio.

Acknowledgment

Our heartfelt appreciation goes to professors K. Kimura and Y. Takagiwa (Grad. School of Frontier Sci., Tokyo Univ.) for enormous support on specimen preparation.